\documentclass[journal]{IEEEtran}

\usepackage{latexsym}
\usepackage{amsfonts}
\usepackage{amsbsy}
\usepackage{amsmath,amssymb}
\usepackage{times}
\usepackage{graphicx}
\usepackage{enumerate}
\usepackage[usenames]{color}
\usepackage[dvips]{pstcol}
\usepackage{epstopdf}
\usepackage{amsmath}
\usepackage{subfig}
\usepackage{bm}
\usepackage{booktabs}
\usepackage{cite}
\usepackage{color}
\usepackage{xcolor}
\usepackage{algorithm}
\usepackage{algorithmicx}
\usepackage{algpseudocode}
\usepackage{setspace}
\usepackage{bm}
\usepackage{tabularx}
\usepackage{amstext,epsfig,psfrag,graphicx,cite,footnote}
\usepackage{siunitx}
\usepackage{mathrsfs}
\usepackage{booktabs}
\usepackage{multirow}
\usepackage{stfloats}
\ifCLASSINFOpdf
\else
\fi
\begin{document}
%
\title{ADMM-based Decoder for Binary Linear Codes Aided by Deep Learning}
%
%
%

\author{Yi Wei, Ming-Min Zhao, Min-Jian Zhao, and Ming Lei \vspace{-0.5cm}

\thanks{The authors are with College of Information Science and Electronic Engineering, Zhejiang University, Hangzhou 310027, China
(email: \{21731133, zmmblack, mjzhao, lm1029\}@zju.edu.cn).
}

}
\IEEEpeerreviewmaketitle

\vspace{-1em}

\maketitle


\begin{abstract}
\color{black}{
Inspired by the recent advances in deep learning (DL), this work presents a deep neural network aided decoding algorithm for binary linear codes. Based on  the
 concept of \emph{deep unfolding}, we design a decoding network by unfolding the alternating direction
method of multipliers (ADMM)-penalized decoder. In addition, we propose two improved versions of the proposed network. The first one transforms the penalty parameter into a set of iteration-dependent ones, and the second one adopts a specially designed penalty function, which is based on a piecewise linear function with adjustable slopes. Numerical results show that the resulting DL-aided decoders outperform the original ADMM-penalized decoder for various low density parity check (LDPC) codes with similar computational complexity.}
\end{abstract}

\begin{IEEEkeywords}
ADMM, binary linear codes, channel decoding, deep learning, deep unfolding.
\end{IEEEkeywords}

\vspace{-0.3cm}
\section{Introduction}
 The linear programming (LP) decoder \cite{1397933}, 
  which is based on LP relaxation of the original maximum likelihood (ML) decoding problem, is {one of many important decoding techniques} for binary linear codes. Due to its strong theoretical guarantees on decoding performance, the LP decoder {has been extensively studied in the literature}, especially for decoding low-density parity-check (LDPC) codes \cite{1397933,4455768}. 
  However,  compared with the classical belief propagation (BP) decoder, the LP decoder has higher computational complexity and poorer error-correcting performance in the low  signal-to-noise-ratio (SNR) region.

 In order to address the above drawbacks, alternating direction method of multipliers (ADMM) has recently been used to solve the LP decoding problem \cite{LiuIT2013,ZhangIT2012,Wei2018,7456284,7740920,JiaoCL2015}. Specifically, the work \cite{LiuIT2013} first presented the ADMM formulation of the LP decoding problem by exploiting the geometry of the LP decoding constraints. The works \cite{ZhangIT2012} and \cite{Wei2018} further reduced the computational complexity of the ADMM decoder. The authors in \cite{7456284} improved the error-correcting performance {through} an ADMM-penalized decoder, where the idea is to make pseudocodewords more costly by adding various  penalty terms to the objective function. Moreover, in \cite{7740920}, the ADMM-penalized decoder was further improved by using piecewise penalty functions, and for irregular LDPC codes, the work \cite{JiaoCL2015} proposed to  modify the penalty term and assign different penalty parameters for variable nodes with different degrees.

Recent advances in deep learning (DL) provide a new direction to tackle tough signal processing tasks in communication systems, such as channel estimation \cite{8715473}, MIMO detection \cite{LcgNet2019} and channel coding \cite{7926071,8242643,SparseNN}. 
 For channel coding, the work \cite{7926071} proposed to use a fully connected neural network and showed that the performance of the network approaches that of the ML decoder for very small block codes.
Then, in \cite{8242643}, the authors proposed to employ the recurrent neural network (RNN) to improve the decoding performance, or alternatively reduce the computational complexity, of a close
to optimal decoder of short BCH codes. {The work \cite{SparseNN} converted the message-passing graph of polar codes into a conventional sparse Tanner graph and proposed  a sparse neural network decoder for
polar codes.}

In this work, we propose to integrate \emph{deep unfolding} {\cite{DeepUnfolding1}} with the ADMM-penalized decoder to improve the decoding performance of  binary linear codes. {This is the first work to construct a deep network by unrolling ADMM-based decoders for binary linear codes.} Different from some classical DL techniques such as the fully connected neural network and the convolutional neural network, which essentially operate as a black box, deep unfolding can make full use of the inherent mechanism of the problem itself and utilize multiple training data to improve the performance with lower training complexity {\cite{DeepUnfolding2}}.  Based on the ADMM-penalized decoder with the cascaded reformulation of the parity check (PC) constraints \cite{4455768}, we propose to construct a learnable ADMM decoding network (LADN) by unfolding the corresponding ADMM iterations, i.e., each stage of LADN can be viewed as one ADMM iteration with some additional adjustable parameters. By following the prototype of LADN, two improved versions are further proposed, which are referred to as {LADN-I and LADN-P}, respectively. In {LADN-I}, we propose to transform the original scalar penalty parameter into a series of iteration-dependent parameters, which can improve the convergence {to reduce the number of iterations} and make performance less dependent on the initial choice of the penalty parameter. Moreover, in {LADN-P}, a specially designed penalty function, i.e., a piecewise linear function with adjustable slopes, is introduced into the proposed LADN, in order to punish pseudocodewords more effectively and increase the decoding performance. {Essentially, we provide a deep learning-based method to obtain a good set of parameters and penalty functions in the ADMM-penalized decoders.} Simulation results demonstrate that the proposed decoders are able to outperform the plain ADMM-penalized decoders with a similar computational complexity.
\vspace{-0.7cm}
\section{Problem Formulation}
\subsection{ML Decoding Problem}
We consider binary linear codes $\mathcal{C}$ of length $N$, each specified by an $M \times N$ PC matrix $\mathbf{H}$.  Throughout this letter, we let $\mathcal{I} \triangleq \{1,\cdots,N\}$ and $\mathcal{J} \triangleq \{1,\cdots,M\}$ denote the sets of variables nodes and check nodes of $\mathcal{C}$, respectively, and let $d_j$ represent the degree of check node $j$. We focus on memoryless binary-input symmetric-output channels, $\mathbf{x}=\{x_i = \{0,1\},i \in \mathcal{I}\}$ is the codeword transmitted over the considered channel and $\mathbf{y}$ is the
received signal.  Then, the ML decoding problem can be formulated
as follows:
%
\begin{subequations} \label{1}
\small
\begin{align}
&\min \limits_{\mathbf{x}} \; \mathbf{v}^T\mathbf{x}\\
&\textrm{s.t.}\; \left[ \sum\nolimits_{i=1}^N H_{ji}x_i\right]_2 = 0,\;j \in \mathcal{J}, \label{parity-check} \\
& \mathbf{x} \in \{0,1\}^{N\times1},\;i \in \mathcal{I}, \label{dicrete_cons}
\end{align}
\end{subequations}

\vspace{-0.2cm}
\noindent where $[\cdot]_2$ denotes the {modulo-2} operator, and $\mathbf{v}=[v_1,\cdots,v_N]^T \in \mathbb{R}^{N\times 1}$ represents the log-likelihood ratio (LLR) vector whose $i$-th element is defined as

\vspace{-0.2cm}
\begin{equation} \label{llr}
v_i = \log \left( \frac{\textrm{Pr}(y_i | x_i = 0)}{\textrm{Pr}(y_i | x_i = 1)} \right),\;i \in \mathcal{I}.
\end{equation}

\vspace{-0.15cm}
\noindent Particularly, $v_i$ can also be viewed as the cost of decoding $x_i = 1$.

Note that the difficulty of solving problem \eqref{1} lies in the PC constraints \eqref{parity-check} and the discrete constraints \eqref{dicrete_cons}. In the next subsection, we will review
the idea of the cascaded reformulation of PC constraints proposed in \cite{4455768} and present the resulting equivalent form of problem \eqref{1}.
\vspace{-0.4cm}
\subsection{Cascaded Formulation of PC Constraints}
 {The key to address the PC constraints in [3] is to decompose the high-degree check nodes into some low-degree ones by introducing auxiliary variables and then recursively employing the three-variable PC transformation, i.e., $[x_1+x_2+x_3]_2 = 0,x_i\in\{0,1\}, i\in\{1,2,3\}$  is transformed into $ \bm{Tx}\preceq \bm{t}, \bm{x}\in \{0,1\}^3$, where

 \vspace{-0.5cm}
\begin{equation}
  \bm{x}=[x_1,x_2,x_3]^T,\bm{t}=[0,0,0,2]^T,
  \begin{array}{l}
\bm{T} = \left[\begin{matrix} 1&-1&-1 \\ -1&1&-1 \\ -1&-1&1 \\ 1 & 1 & 1 \end{matrix}\right].
\end{array}
\end{equation}

\vspace{-0.2cm}
\noindent In order to express the PC constraints in a more compact form, we define
\begin{equation}\label{ut}
\begin{aligned}
&\mathbf{u} = [{\mathbf{x}}^T,{\tilde{\mathbf{x}}}^T]^T \in \{0,1\}^{{(N+\Gamma_a)}\times1}, \\
&\mathbf{t} = [0,0,0,2]^T, \mathbf{b} = \mathbf{1}_{\Gamma_c \times 1} \otimes \mathbf{t},
\mathbf{q} = [\mathbf{v};\mathbf{0}_{\Gamma_a \times 1}],\\
&\mathbf{A} = [\mathbf{T} \mathbf{Q}_1;\cdots;\mathbf{T} \mathbf{Q}_{\tau};\cdots;\mathbf{T} \mathbf{Q}_{\Gamma_c}] \in \{0,\pm 1\}^{4\Gamma_c \times (N+\Gamma_a)},
\end{aligned}
\end{equation}
where $\Gamma_a = \sum_{j=1}^M (d_j-3)$ and $\Gamma_c = \sum_{j=1}^M (d_j-2)$ represent the total numbers of auxiliary variables and three-variable PC equations,  $\tilde{\mathbf{x}}=[{\tilde{x}}_1,{\tilde{x}}_2,\cdots,{\tilde{x}}_{\Gamma_a}]^T$, and $\mathbf{Q}_{\tau}\in \{0,1\}^{(N+\Gamma_a)\times1}$ denotes a selection matrix that chooses the corresponding variables in $\mathbf{u}$ which are involved in the $\tau$-th three-variable PC equation. Then, we can see that the ML decoding problem (1) is equivalent to the following linear integer programming problem:

\vspace{-0.5cm}
\begin{subequations} \label{ML_problem_Reformulation}
\begin{align}
&\min\limits_{\mathbf{u}} \; \mathbf{q}^T\mathbf{u}\label{objective}\\
&\textrm{s.t.}\; \mathbf{A} \mathbf{u} -\mathbf{b} \preceq \mathbf{0}, \label{PC_constraint}\\
&\mathbf{u} \in \{0,1\}^{(N+\Gamma_a)\times1}.\label{discrete_constraint}
\end{align}
\end{subequations}}
\vspace{-1cm}
\section{Learned ADMM decoder}
In this section, we first review the ADMM-penalized decoder to address problem \eqref{ML_problem_Reformulation}, and then  we present a detailed description of the proposed LADN and its improved versions. Finally, we provide the loss function of the proposed networks, which is essential to achieve better decoding performance.
\subsection{ADMM-Penalized Decoder}
 The essence of the ADMM-penalized decoder is the introduction of a penalty term to the linear objective of the LP decoding problem, with the intent of suppressing the pseudocodewords. In order to put problem \eqref{ML_problem_Reformulation} in the standard ADMM framework, an auxiliary variable $\mathbf{z}$ is first added to constraint \eqref{PC_constraint}, and consequently, problem  \eqref{ML_problem_Reformulation} can be equivalently formulated
as the following optimization problem:

\vspace{-0.4cm}
\begin{subequations} \label{ML_problem_Reformulation1} 
\begin{small}
\begin{flalign}
&\min \limits_{\mathbf{u},\mathbf{z}} \; \mathbf{q}^T\mathbf{u}\label{objective}\\
&\textrm{s.t.}\; \mathbf{A} \mathbf{u} + \mathbf{z} = \mathbf{b}, \label{PC_constraint1}\\
&\mathbf{u} \in \{0,1\}^{(N+\Gamma_a)\times1}, \mathbf{z}\in {\mathbb{R}}_{+}^{4{\Gamma}_a\times 1}.\label{discrete_constraint}
\end{flalign}
\end{small}
\end{subequations}

\vspace{-0.6cm}
\noindent{Next, the discrete constraint \eqref{discrete_constraint} is relaxed to  $\mathbf{u}\in [0,1]^{(\Gamma_a+n)\times 1}$  and we penalize the pseudocodewords using penalty functions that make integral solutions more favorable than fractional solutions, which leads to the following problem:}
\begin{equation} \label{PenalOP1}
\begin{small}
\begin{aligned}
&\min \limits_{\mathbf{u},\mathbf{z}} \; \mathbf{q}^T\mathbf{u}+\sum_ig(u_i)\\
&\textrm{s.t.}\; \mathbf{A} \mathbf{u} + \mathbf{z} = \mathbf{b},\\
&\mathbf{u} \in [0,1]^{(N+\Gamma_a)\times1}, \mathbf{z}\in {\mathbb{R}}_{+}^{4{\Gamma}_a\times 1}.
\end{aligned}
\end{small}
\end{equation}
In \eqref{PenalOP1}, $g(\cdot):[0,1]\rightarrow \mathbb{R}\cup \{\pm\infty\}$ is the introduced penalty function, e.g.,  the L1 or L2 function used in \cite{7456284}.

The augmented Lagrangian function of problem \eqref{PenalOP1} can be formulated as
\begin{equation}
\begin{small}
\begin{aligned}
{L_{\mu}(\mathbf{u},\mathbf{z},\mathbf{y})}&=\mathbf{q}^T\mathbf{u}+\sum_ig(u_i)\\
&+{\mathbf{y}}^T(\mathbf{Au}+\mathbf{z}-\mathbf{b})+\frac{\mu}{2}||\mathbf{Au}+\mathbf{z}-\mathbf{b}||_2^2,
\end{aligned}
\end{small}
\end{equation}

\vspace{-0.2cm}
\noindent where $\mathbf{y}\in {\mathbb{R}}^{4\Gamma_c\times 1}$ denotes the Lagrangian multiplier and $\mu$ represents a positive penalty parameter. Then, the iterations of ADMM can be written as

\vspace{-0.4cm}
\begin{subequations}
\begin{small}
\begin{align}
  &{\mathbf{u}}^{k+1} = \mathop {\arg\min}\limits_{\mathbf{u}\in [0,1]^{(N+{\Gamma}_a)\times1}}L_{\mu}(\mathbf{u},{\mathbf{z}}^k,{\mathbf{y}}^k),\label{ADMM1}\\
  &{\mathbf{z}}^{k+1} = \mathop {\arg\min}\limits_{\mathbf{z}\in {\mathbb{R}}_+^{{\Gamma}_c\times1}}L_{\mu}({\mathbf{u}}^{k+1},{\mathbf{z}},{\mathbf{y}}^k),\label{ADMM2}\\
  &{\mathbf{y}}^{k+1}= {\mathbf{y}}^k+\mu(\mathbf{A}{\mathbf{u}}^{k+1}+{\mathbf{z}}^{k+1}-\mathbf{b}).\label{y_update}
\end{align}
\end{small}
\end{subequations}

\vspace{-0.5cm}
 \noindent Since $\mathbf{A}$ is orthogonal in columns, we can see that ${\mathbf{A}}^T\mathbf{A}$ is a diagonal matrix and the variables in (\ref{ADMM1}) are separable. Therefore, step (\ref{ADMM1}) can be conducted by solving the following $N+{\Gamma}_a$ parallel subproblems:

 \vspace{-0.4cm}
\begin{subequations}\label{a_op}
\begin{small}
\begin{align}
&\mathop{\min}\limits_{\mathbf{u}}\frac{1}{2}\mu e_i u_i^2+g(u_i)+(q_i+{\mathbf{a}}_i^T(\mathbf{y}+\mu({\mathbf{z}}^k-\mathbf{b})))u_i,\label{a_op1}\\
&\textrm{s.t.}\; u_i\in [0,1],i \in \mathcal{I},
\end{align}
\end{small}
\end{subequations}

\vspace{-0.5cm}
 \noindent where ${\mathbf{a}}_i$ denotes the $i$-th column of ${\mathbf{A}}$, $\mathbf{e}=\textrm{diag}({\mathbf{A}}^T\mathbf{A})=[e_1,\cdots,e_{N+{\Gamma}_a}]$.
With a well-designed penalty function $g(\cdot)$, (\ref{a_op1}) is guaranteed to be convex and the optimal solution of problem (\ref{a_op}) can be easily obtained  by setting the gradient of  (\ref{a_op1}) w.r.t. $u_i$ to zero and then projecting the resulting solution to the interval [0,1].
Similarly, the optimal solution of problem (\ref{ADMM2}) can be obtained by
\vspace{-0.2cm}
\begin{equation}\label{z_update}
\begin{small}
{\mathbf{z}}^{k+1}={\Pi}_{[0,+\infty]^{4{\Gamma}_c}}\left(\mathbf{b}-{\mathbf{Au}}^{k+1}-\frac{{\mathbf{y}}^k}{\mu}\right),
\end{small}
\end{equation}
\vspace{-0.3cm}

\noindent where ${\Pi}_{[0,\infty]}(\cdot)$ denotes the Euclidean projection operator which projects the resulting solution to the interval $[0,\infty]$.


To summarize, the ADMM-penalized decoder iterates over three steps,  i.e.,  \eqref{ADMM1}-\eqref{y_update}, and the final estimated codeword $\hat{\mathbf{x}}$ is obtained by $\hat{\mathbf{x}}=\Pi_{\{0,1\}}([u_1,\cdots,u_N])$, where $\Pi_{\{0,1\}}(s)=0$ if $s\textless 0.5$, and $\Pi_{\{0,1\}}(s)=1$ otherwise.
\vspace{-0.3cm}
\subsection{The Proposed LADN}
Unfolding a well-understood iterative algorithm (also known as deep unfolding) is one of the most popular and powerful techniques to build a model-driven DL network. The resulting networks have been shown to outperform their baseline algorithms in many cases, such as the LAMP \cite{Borgerding2016Onsager} , the ADMM-net \cite{Yang2016Deep} and the LcgNet \cite{LcgNet2019}, etc. Based on the aforementioned ADMM-penalized decoder, we construct our LADN by unfolding the iterations of \eqref{ADMM1}-\eqref{y_update} and regarding $\mu$ and the coefficients in $g(\cdot)$ as learnable parameters, {i.e., $\{\alpha,\mu\}$. Note that training a single parameter $\alpha$ or $\mu$ also helps to improve the baseline ADMM L2 decoder, however the performance gain is inferior to that achieved by LADN with two parameters learned jointly.}

For the purpose of illustration, we consider the L2 penalty function here, whose definition is given by
 $g_{L2}(u)=-\frac{\alpha}{2}(u-0.5)^2$,
  where $\alpha$ is the coefficient that controls the slope of $ g_{L2}(\cdot)$.
   Then, the solution of problem \eqref{a_op} can be explicitly obtained by
 \begin{equation}\label{uup}
  u_i^{k+1}= {\Pi}_{[0,1]}\left(\frac{q_i+{\mathbf{a}}_i^T({\mathbf{y}}^k+\mu({\mathbf{z}}^k-\mathbf{b}))+\frac{\alpha}{2}}{-\mu e_i+\alpha}\right).
 \end{equation}
 For convenience, the ADMM-penalized decoder with the L2 penalty function is referred to as ADMM L2 decoder in the following.
\begin{figure}[t]
\vspace{-0.2cm}
\setlength{\belowcaptionskip}{-0.6cm}
\renewcommand{\captionfont}{\small}
\centering
\includegraphics[scale=.28]{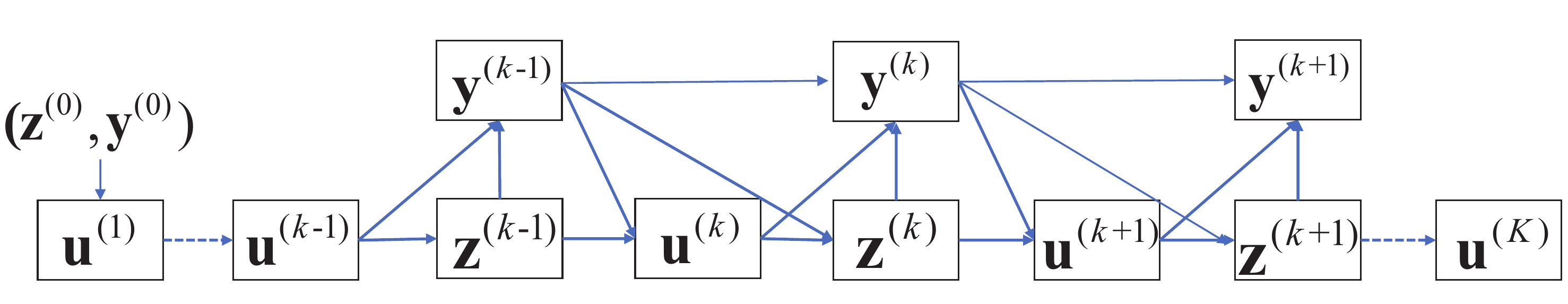}
\caption{The structure of the proposed LADN.}
\label{LADMM}
\normalsize
\end{figure}

The proposed LADN is defined over a data flow graph based on the ADMM iterations, which is depicted in Fig. \ref{LADMM}. The nodes in the graph correspond to different operations in the ADMM L2 decoder, and the directed edges represent the data flows between these operations.  LADN consists
of $K$ stages each with the same structure, and the $k$-th stage corresponds
to the $k$-th iteration of the ADMM-penalized decoder. Each stage includes three nodes, i.e., the $\mathbf{u}$-node, the $\mathbf{z}$-node and the $\mathbf{y}$-node, which correspond to the updating steps in \eqref{ADMM1}-\eqref{y_update}. Let $({\mathbf{u}}^{(k)},{\mathbf{z}}^{(k)},{\mathbf{y}}^{(k)})$ denote the outputs of these nodes in the $k$-th stage, the detailed steps when calculating $({\mathbf{u}}^{(k)},{\mathbf{z}}^{(k)},{\mathbf{y}}^{(k)})$ can be expressed as (also shown in Fig. 2)

\vspace{-0.4cm}
\begin{subequations}\label{DecNet}
\begin{small}
\begin{align}
  &{\mathbf{u}}^{(k+1)}= {\Pi}_{[0,1]}\left(\bm{\eta}\odot(\mathbf{q}+{\mathbf{A}}^T({\mathbf{y}}^{(k)}+\mu({\mathbf{z}}^{(k)}-\mathbf{b}))+\frac{\alpha}{2})\right),\label{u1}\\
&{\mathbf{z}}^{(k+1)}={\textrm{ReLU}}\left(\mathbf{b}-{\mathbf{Au}}^{(k+1)}-{{\mathbf{y}}^{(k)}}/{\mu}\right),\\
&{\mathbf{y}}^{(k+1)}= {\mathbf{y}}^{(k)}+\mu(\mathbf{A}{\mathbf{u}}^{(k+1)}+{\mathbf{z}}^{(k+1)}-\mathbf{b}),
\end{align}
\end{small}
\end{subequations}

\vspace{-0.6cm}
 \noindent where $\bm{\eta}\in {\mathbb{R}}^{N+{\Gamma}_a}$ is the output of the function $ \eta(\mathbf{A};\alpha;\mu) \triangleq \textrm{diag}\left({1}/({\alpha-\mu{\mathbf{A}}^T\mathbf{A}})\right)$, and the symbol $\odot$ represents the {Hadamard} product. Besides, {$\textrm{ReLU}(\cdot)$} denotes the classical active function in the DL field, i.e., {$\textrm{ReLU}(x) = \max(x; 0)$},
which is equivalent to the projection operation ${\Pi}_{[0,\infty]}(\cdot)$ in \eqref{z_update}. The parameters $\{\mu,\alpha\}$ in \eqref{DecNet} are considered as learnable parameters to be trained and the final decoding output of the proposed network, i.e., $\hat{\mathbf{x}}$, can be obtained by $\hat{\mathbf{x}} = \Pi_{\{0,1\}}([u^{(K)}_1,\cdots,u^{(K)}_N])$.

\begin{figure}[t]
\vspace{-0.2cm}
\setlength{\belowcaptionskip}{-0.6cm}
\renewcommand{\captionfont}{\small}
\centering
\includegraphics[scale=.32]{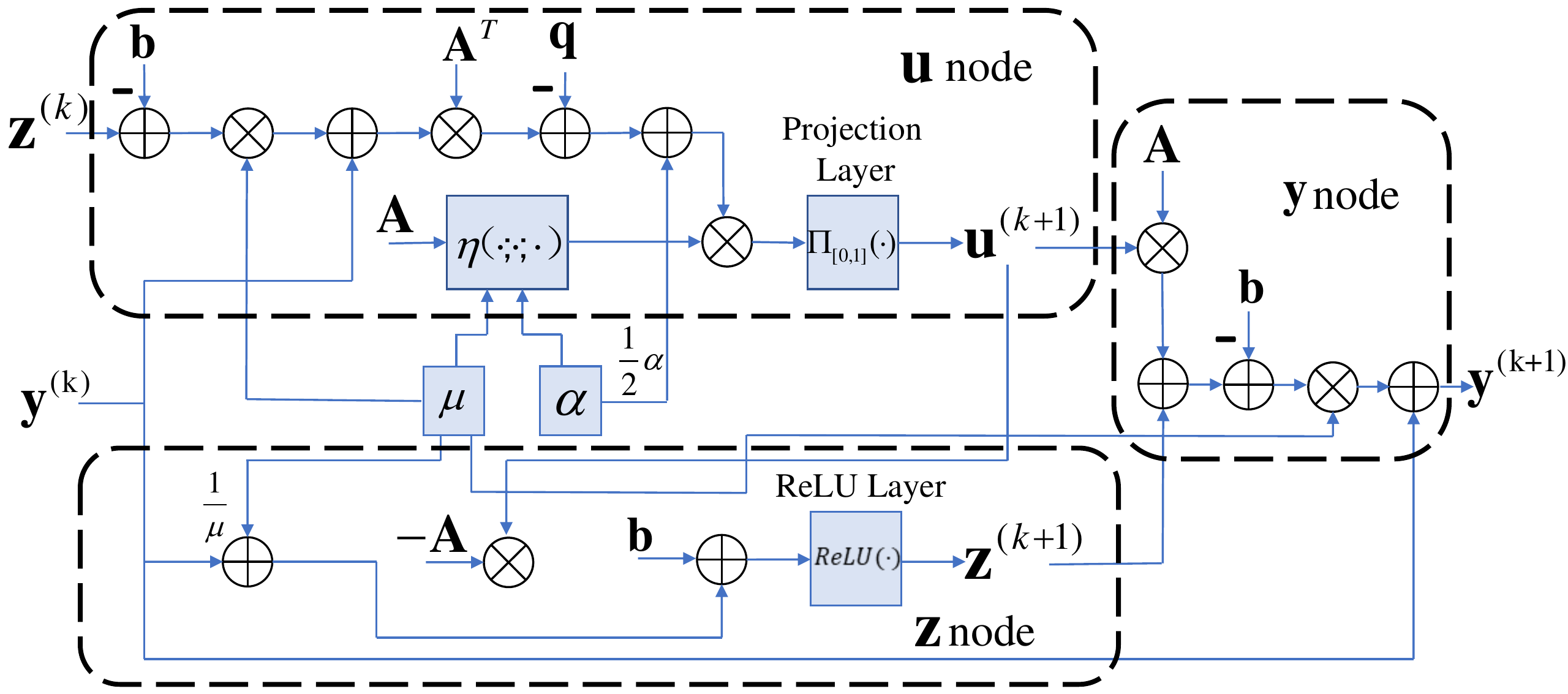}
\caption{The $k$-th stage of LADN with learnable parameters $\alpha$ and $\mu$.}
\label{nodes}
\normalsize
\end{figure}
\vspace{-0.3cm}
\subsection{{LADN-I} with Iteration-Dependent Penalty Parameters}
In order to improve the performance of LADN,  we propose to increase the number of learnable parameters, and the resulting network  is referred to as {LADN-I}. The main idea of this improved network is to transform the penalty parameter $\mu$ into a series of iteration-dependent ones.
 This is based on the intuition that increasing the number of learnable parameters (or network scale) is able to improve the generalization ability of neural networks. Besides, using possibly different penalty parameters
for each iteration/stage can potentially  improve the convergence in practice, as well as make performance less dependent on the initial choice of the penalty parameter. Since the proposed varying penalty decoder includes the conventional fixed  penalty decoder as a special case, we can infer that {LADN-I} incurs no loss of optimality in general.

More specifically, we employ $\bm{\mu}=[\mu_1,\cdots,\mu_K]$ as learnable parameters, where $\mu_k$ denotes the penalty parameter in the $k$-th stage.
With $\{\alpha,\bm{\mu}\}$, the decoding steps in \eqref{DecNet} can be rewritten as follows:

\vspace{-0.5cm}
\begin{subequations}
\begin{small}
\begin{align}
 &{\mathbf{u}}^{(k+1)}= {\Pi}_{[0,1]}\left(\bm{\eta}\odot(\mathbf{q}+{\mathbf{A}}^T({\mathbf{y}}^{(k)}+\mu_k({\mathbf{z}}^{(k)}-\mathbf{b}))+\frac{\alpha}{2})\right),\\
&{\mathbf{z}}^{(k+1)}={\textrm{ReLU}}\left(\mathbf{b}-{\mathbf{Au}}^{(k+1)}-{{\mathbf{y}}^{(k)}}/{\mu_k}\right),\\
&{\mathbf{y}}^{(k+1)}= {\mathbf{y}}^{(k)}+\mu_k(\mathbf{A}{\mathbf{u}}^{(k+1)}+{\mathbf{z}}^{(k+1)}-\mathbf{b}).
\end{align}
\end{small}
\end{subequations}
%
\vspace{-1cm}
\subsection{{LADN-P} with Learnable Penalty Function}
In this subsection, we propose another improved version of  LADN by introducing a novel adjustable penalty function, and the resulting network is named as {LADN-P}. This network is based on the observation that the choice of the penalty function $g(\cdot)$ has a vital impact on the performance of the ADMM-penalized decoder and designing a good penalty function with the aid of DL can potentially improve the decoding performance. According to \cite{7456284}, $g(\cdot)$ should satisfy the following properties: 1) $g(\cdot)$ is an increasing function on $[0, 0.5]$; 2) $g(u)=g(1-u)$ for $x\in [0,1]$; 3) $g(\cdot)$ is differentiable on $[0, 0.5]$; 4) $g(\cdot)$ is such that the solution of the $\mathbf{u}$-update problem \eqref{a_op} is well-defined.

Note that an improved piecewise penalty function was proposed in \cite{7740920} by increasing the slope of the penalty function at the points near $x = 0$ and $x = 1$, and the parameters that controls these slopes are decided by first choosing the possible set of parameters empirically and then simulating the FER performance for all possible combinations of the parameters. The number of pieces can not be large due to the fact that the search process of these parameters is complex and time-consuming.  To address this difficulty, we propose a learnable linear penalty function whose slope parameters can be obtained by training.
Since a  piecewise linear function is able to approximate any nonlinear function when  the number of pieces is large enough, we can learn a flexible penalty function which is able to outperform the conventional L1 and L2 functions. Besides, by resorting to power of DL, the corresponding parameters can be trained from data through back propagation,  instead of  empirically tuning or exhaustive search.

The definition of
the proposed adjustable penalty function is given by

\vspace{-0.5cm}
\begin{equation}\label{glearn}
\small
g_{l}(x)=
\left\{ \begin{array}{ll}
             \phi_lx+\beta_l,  &c_{l-1} \le x \textless c_l, \\
                               & l \in \{1,2,\cdots,L-1\}\\
            \phi_Lx+\beta_L,  &c_{L-1}\le x\le 0.5\\
            -\phi_Lx+\beta_{L+1},  &0.5 \textless  x \le 1-c_{L-1}\\
           -\phi_lx+\beta_{2L+1-l},  &1-c_l \textless  x \le 1-c_{l-1},\\
                                     &l \in \{L-1,L-2,\cdots,1\}
            \end{array}\right.
\end{equation}

\vspace{-0.2cm}
\noindent where $2L$ is the number of pieces, $\{\phi_l\}_{l=1}^{L}$ is the slope set, $\{\beta_l\}_{l=1}^{2L}$ denotes the bias set and $\{c_{l-1}\}_{l=1}^{L}$ contains the predefined positions which are uniformly located within $[0,0.5]$ with $c_0 = 0$.
It can be observed that the proposed penalty function is symmetrical and differentiable {almost} everywhere on the interval [0,1], which meets property 2 and 3 mentioned above. In order to satisfy property 1, we can simply initialize $\{\phi_l\}_{l=1}^L$ to be positive.
Besides, due to the  piecewise linearity of $g_l(x)$, the ${\mathbf{u}}$-update step can be easily derived by  resorting to the first-order optimality condition of problem \eqref{a_op}, i.e.,

\vspace{-0.5cm}
\begin{equation}\label{slope}
\small
u_i^{(k)} = \left\{ \begin{array}{ll}
             \Pi_{[0,1]}\left(\frac{t_i-\phi_1}{h_i}\right),  &t_i \textless \delta_1\\
             \Pi_{[0,1]}\left(\frac{t_i-\phi_l}{h_i}\right),  &\delta_l \le t_i \textless \delta_{l+1}, \\
                                                              & l\in\{2,3,\cdots,L-1\} \\
             \Pi_{[0,1]}\left(\frac{t_i-\phi_L}{h_i}\right),   &\delta_L \le t_i \textless \frac{h_i}{2}\\
             \Pi_{[0,1]}\left(\frac{t_i+\phi_L}{h_i}\right),   &\frac{h_i}{2}  \le t_i \textless \psi_L \\
             \Pi_{[0,1]}\left(\frac{t_i+\phi_l}{h_i}\right),   &\psi_l  \le t_i \textless \psi_{l-1},\\
                                                               & l\in \{L-1,L-2,\cdots,2\}\\
             \Pi_{[0,1]}\left(\frac{t_i+\phi_1}{h_i}\right),  &t_i \geq \psi_{1}
            \end{array}\right.
\end{equation}

\vspace{-0.2cm}
\noindent where $\delta_l=c_{l}h_i+\phi_{l+1}$, $\psi_l=(1-c_l)h_i+\phi_{l+1}$, $t_i= q_i+{\mathbf{a}}_i^T( \mu(\mathbf{b}-\mathbf{z})-\mathbf{y})$ and $h_i = \mu {e}_i$.  Therefore, \eqref{glearn} also satisfies property 4. {It can be further observed that the slopes of $g_l(\cdot)$, i.e., $\{\phi_l\}_{l=1}^L$,  are required in the $\mathbf{u}$-update step and the biases $\{\beta_l\}_{l=1}^{2L}$ are not included in \eqref{slope}, thus $\{\phi_l\}_{l=1}^L$ are the only training parameters.}
\vspace{-0.5cm}
\subsection{Loss Function}
Let $\{{\mathbf{v}}_p,{\mathbf{x}}_p\}_{p=1}^{P}$ denote the set of training samples with size $P$, where the LLR vector ${\mathbf{v}}_p$ and the transmit codeword ${\mathbf{x}}_p$ are viewed as the $p$-th feature and label, respectively. After accepting ${\mathbf{v}}_p$ as input, the proposed networks are expected to predict ${\mathbf{x}}_p$ that corresponds to this ${\mathbf{v}}_p$. In the following, we let ${\mathcal{F}}_{\textrm{LADN}}(\cdot)$ denote the underlying mapping performed by the proposed networks, which satisfies
$\hat{\mathbf{x}}= {\mathcal{F}}_{\textrm{LADN}}(\mathbf{v};\bm{\Theta})$ and ${\bm{\Theta}}$ denotes the collection of learnable parameters in the proposed networks, e.g., $\{\alpha,{\mu}\}$ in LADN, $\{\alpha,\bm{\mu}\}$ in {LADN-I} and $\{\{\phi_l\}_{l=1}^L,{\mu}\}$ in {LADN-P}.

All learnable parameters in the proposed networks can be optimized by utilizing  the training samples $({\mathbf{v}}_p,{\mathbf{x}}_p)_{p=1}^{P}$  to minimize a certain loss function. For the purpose of improving the decoding performance, we design a novel loss function based on the mean square error (MSE) criterion, whose definition is given by
\vspace{-0.25cm}
\begin{equation}\label{loss}
\begin{small}
\begin{aligned}
{\mathcal{L}}({\bm{\Theta}})&=\frac{1}{P}\sum\limits_{p = 1}^P\Big({\sigma}||{\mathbf{Au}}_p^{(K)}+{\mathbf{z}}_p^{(K)}-\mathbf{b}||_2^2\\
&+ (1-\sigma)|| {\mathcal{F}}_{\textrm{LADN}}({\mathbf{v}}_p;\bm{\Theta})-{\mathbf{x}}_p||_2^2\Big).
\end{aligned}
\end{small}
\end{equation}
As shown in (\ref{loss}), ${\mathcal{L}}(\cdot)$ is composed of the weighted sum of an unsupervised term and a supervised term, i.e.,  $||{\mathbf{Au}}_p^{(K)}+{\mathbf{z}}_p^{(K)}-\mathbf{b}||_2^2$ and {$|| {\mathcal{F}}_{\textrm{LADN}}({\mathbf{v}}_p;\bm{\Theta})-{\mathbf{x}}_p||_2^2$}, with $\sigma$ being the weighting factor. More specifically, the unsupervised term measures the power of the residual between ${\mathbf{Au}}^{(k)}+{\mathbf{z}}^{(k)}$ and $\mathbf{b}$. The decoding process is considered to be completed when this residual is sufficiently small, i.e., $||{\mathbf{Au}}^{(k)}+{\mathbf{z}}^{(k)}-\mathbf{b}||_2^2 \le \epsilon$, which also means that constraint \eqref{PC_constraint1} is nearly satisfied. Therefore, employing this residual as part of the loss function helps to accelerate the decoding process. Besides, the supervised term aims to minimize the distance between the network output ${\mathcal{F}}_{\textrm{LADN}}(\cdot)$ and the true transmit codeword ${\mathbf{x}}_p$, which is beneficial for improving decoding accuracy. {Note that  in order to learn a decoder which is effective under various numbers of iterations,  the proposed loss function  takes the outputs of
all layers into consideration. }
\vspace{-0.3cm}
\section{Simulation Results}
In this section, computer simulations are carried out to evaluate the performance of the  proposed LADN and its improved versions. All networks are implemented in Python using the TensorFlow library with the Adam optimizer \cite{ADAM}. In the simulations, we focus on additive white Gaussian noise (AWGN) channel with binary phase shift keying (BPSK) modulation.
The considered binary linear codes are [96, 48] MacKay 96.33.964 LDPC code $\mathcal{C}_1$ and [128, 64] CCSDS LDPC code $\mathcal{C}_2$ \cite{channelcodes}.

 It is noteworthy that the training SNR plays an important role in generating the training samples.
 If the training SNR is set too high, very few decoding errors exist and the networks may fail to learn the underlying error patterns. However, if the SNR is too low, only few transmitted codewords can be successively decoded and this will prevent the proposed networks from learning the effective decoding mechanism. In this work, we set the training SNR to 2dB, which is obtained by cross-validation \cite{crvali1995}.
The training and validation data sets contain $4\times 10^4$ and {$10^4$} samples, {and the message bits can be all-zeros or randomly generated.\footnote{{Note that although we have verified by simulations that all-zero codewords can also be used for training, a rigorous proof on whether the proposed decoders satisfy the all-zero assumption \cite{7456284} (i.e., the symmetry condition) remains an open problem, which is out of the scope of this paper.}}}  The hyper-parameter $\sigma$ in ${\mathcal{L}}(\cdot)$ is set to 0.3 and 0.9 for $\mathcal{C}_1$ and $\mathcal{C}_2$, respectively, which are chosen by cross-validation \cite{crvali1995}. In {LADN-P}, we set the number of pieces in \eqref{glearn} to $2L=10$. We employ a decaying learning rate which is set to 0.001 initially and then reduce it by half every epoch. The training process is terminated when the average validation loss stops decreasing. Besides, it is important to note that in the offline training phase, the number of stages is fixed to $50$ and $70$ for ${\mathcal{C}}_1$ and  ${\mathcal{C}}_2$ codes, respectively. {According to (9c), (11) and (12),  the total computational complexity of the ADMM L2 decoder in each iteration is roughly $\mathcal{O}(N + \Gamma a)$ real multiplications + $\mathcal{O}(10(N + \Gamma a)-1)$ real applications + 2 real divisions.  Since LADN (LADN I) finally perform as the ADMM L2 decoder loaded with learned parameters $\{\alpha,\mu\}$ ($\{\alpha,\bm{\mu}\}$),
its computational complexity is the  same as the ADMM L2 decoder, which is lower than that of the ML decoder, i.e., $\mathcal{O}(2^N)$.}

\begin{figure}[t]
\vspace{-0.6cm}
\setlength{\belowcaptionskip}{-0.5cm}
\renewcommand{\captionfont}{\small}
\centering
\includegraphics[scale=.39]{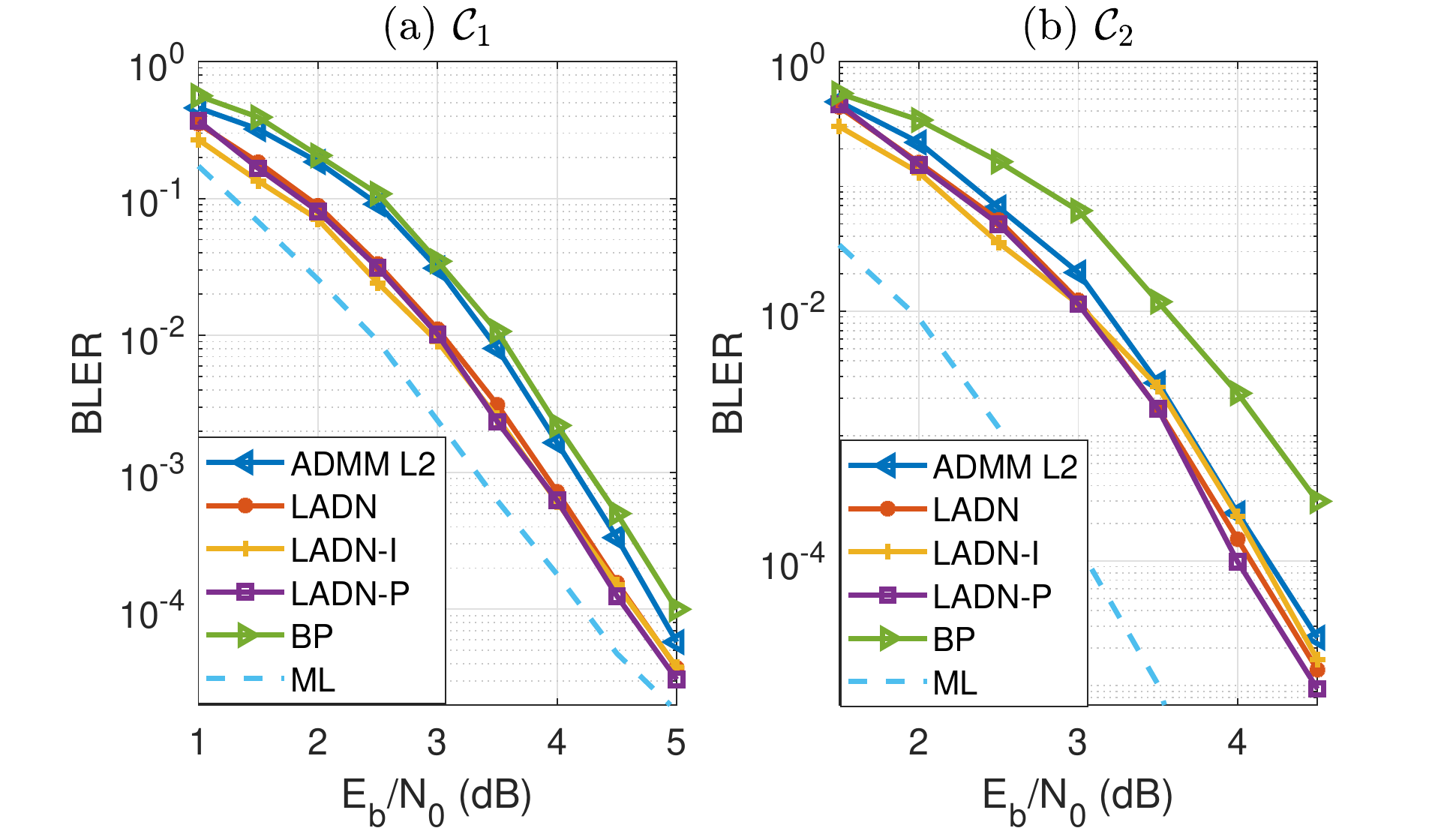}
\caption{BLER performance comparison of $\mathcal{C}_1$ and $\mathcal{C}_2$ codes.}
\label{BLER}
\normalsize
\end{figure}

Fig. \ref{BLER} demonstrates the block error rate (BLER) performance comparison of the BP decoder, the ADMM L2 decoder and the proposed decoding networks. For the ADMM L2 decoder, $\alpha$ and $\mu$ are set to 1 and 1.2, respectively. In all the curves, we collect at least 100 block errors for all data points.  It can be observed that for both codes, our proposed networks show better BLER performance over the original ADMM L2 decoder at both low and high SNR regions. The proposed {LADN-I} achieves the best BLER performance at the low SNR region, and {LADN-P} outperforms the other counterparts when SNR is high. {Note that with the increasing of the code length, the training complexity also increases and it remains to be investigated whether the proposed method can still achieve a noticeable performance gain for longer codes.  }

\begin{figure}[t]
\vspace{-0.6cm}
\setlength{\belowcaptionskip}{-0.6cm}
\renewcommand{\captionfont}{\small}
\centering
\includegraphics[scale=.53]{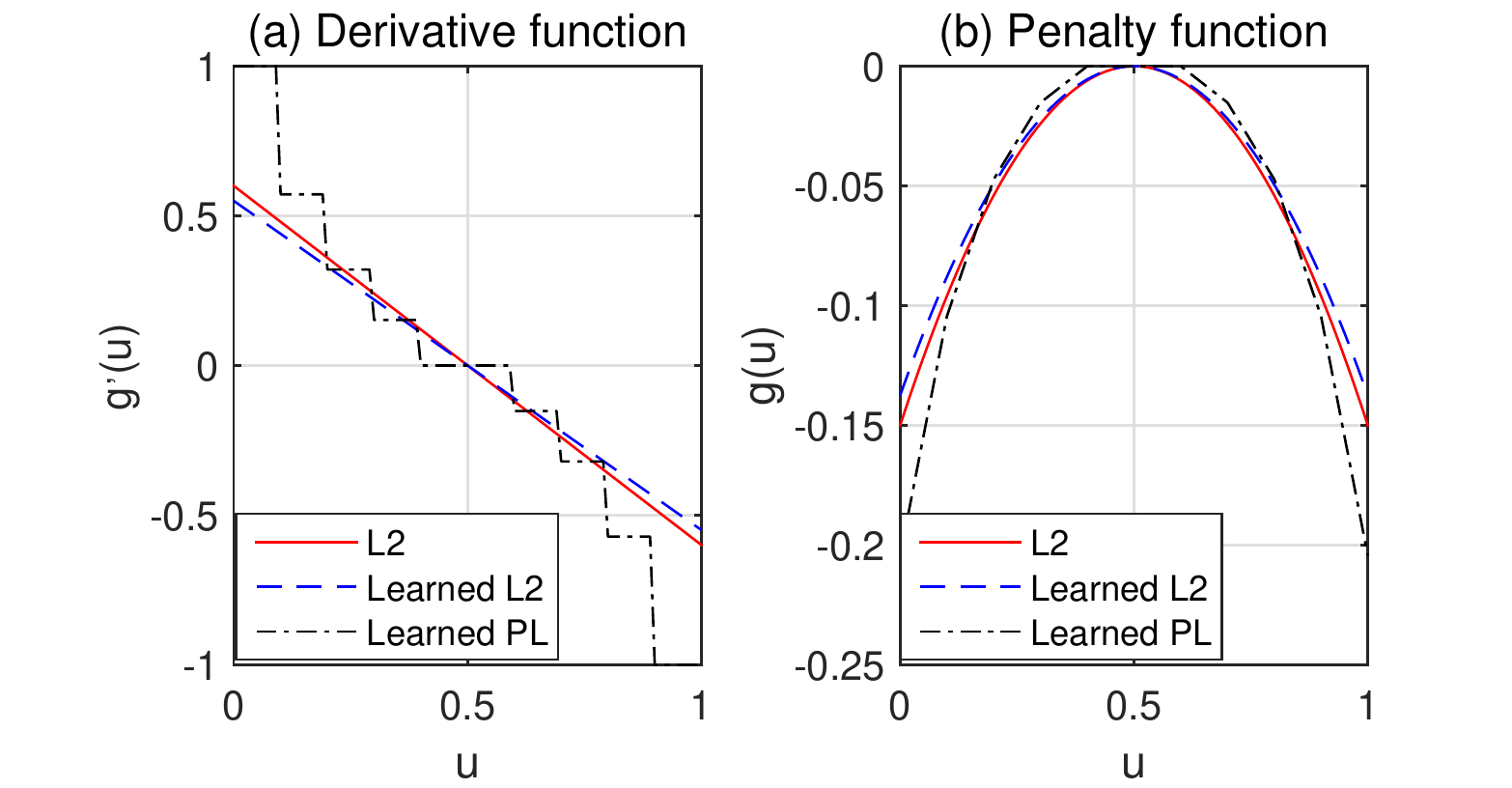}
\caption{ The learned penalty functions and their corresponding derivative functions for $\mathcal{C}_1$ code.}
\label{penal}
\normalsize
\end{figure}

Moreover, in Fig. \ref{penal}, we show the curves of the penalty functions employed in the considered decoders (for ${\mathcal{C}}_1$ code) to illustrate their properties, where \emph{L2}, \emph{Learned L2} and \emph{Learned PL} denote the L2 penalty function in the ADMM L2 decoder, the L2 penalty function with learned parameter $\alpha$ in LADN and the adjustable piecewise linear penalty function in {LADN-P}, respectively. {Note that the {absolute values of the} slopes of the penalty function $|g_h'(u)|$  at the points near $u=0.5$ should be small, due to the fact that a larger slope may prevent the ADMM-penalized decoder
from forcing the variables with values near 0.5 to 0 or 1. On the contrary, the values of $|g_h'(u)|$ at the points near $u = 0$ or 1 should be large. Therefore, empirically, higher-order polynomial functions, such as  $g_h(u)= -\frac{\alpha}{2}(u-0.5)^4$, are better than the \emph{L2} penalty function. However, the solution of problem (10) is not well-defined if these higher-order functions are employed as penalty functions. It can be observed from Fig. 3 (a) that the \emph{Learned PL} function has the largest {absolute values of the} slopes when $u=0$ or 1 and the smallest ones when $u=0.5$ compared with those of \emph{L2} and \emph{Learned L2}.} Furthermore, from Fig. \ref{penal} (b), we can see that the curves of the \emph{Learned PL} function is similar to a higher-order (larger than 2) polynomial function, however the solution of the $\mathbf{u}$-update problem \eqref{a_op} in this case is well defined since the \emph{Learned PL} function is composed of linear functions.
\vspace{-0.4cm}
\section{conclusion}
In this work, we adopted the deep unfolding technique to improve the performance of the ADMM-penalized decoder for binary linear codes. The proposed decoding network is essentially obtained by unfolding the iterations of the ADMM-penalized decoder and transforming some preset parameters into learnable ones.   Furthermore, we presented two improved versions of the proposed network by transforming the penalty parameter into a series of iteration-dependent parameters and introducing a specially designed adjustable penalty function. {Numerical results were presented to show that the proposed networks outperform the plain ADMM-penalized decoder with  similar complexity.}
\ifCLASSOPTIONcaptionsoff
  \newpage
\fi



%
%
%
\vspace{-0.3cm}
\bibliography{ADMM}

\begin{thebibliography}{10}
\providecommand{\url}[1]{#1}
\csname url@samestyle\endcsname
\providecommand{\newblock}{\relax}
\providecommand{\bibinfo}[2]{#2}
\providecommand{\BIBentrySTDinterwordspacing}{\spaceskip=0pt\relax}
\providecommand{\BIBentryALTinterwordstretchfactor}{4}
\providecommand{\BIBentryALTinterwordspacing}{\spaceskip=\fontdimen2\font plus
\BIBentryALTinterwordstretchfactor\fontdimen3\font minus
  \fontdimen4\font\relax}
\providecommand{\BIBforeignlanguage}[2]{{%
\expandafter\ifx\csname l@#1\endcsname\relax
\typeout{** WARNING: IEEEtran.bst: No hyphenation pattern has been}%
\typeout{** loaded for the language `#1'. Using the pattern for}%
\typeout{** the default language instead.}%
\else
\language=\csname l@#1\endcsname
\fi
#2}}
\providecommand{\BIBdecl}{\relax}
\BIBdecl

\bibitem{1397933}
J.~{Feldman}, M.~J. {Wainwright}, and D.~R. {Karger}, ``Using linear
  programming to decode binary linear codes,'' \emph{IEEE Trans. Inf. Theory},
  vol.~51, no.~3, pp. 954--972, Mar. 2005.

\bibitem{4455768}
K.~{Yang}, X.~{Wang}, and J.~{Feldman}, ``A new linear programming approach to
  decoding linear block codes,'' \emph{IEEE Trans. Inf. Theory}, vol.~54,
  no.~3, pp. 1061--1072, Mar. 2008.

\bibitem{LiuIT2013}
S.~Barman, X.~Liu, S.~C. Draper, and B.~Recht, ``Decomposition methods for
  large scale {LP} decoding,'' \emph{IEEE Trans. Inf. Theory}, vol.~59, no.~12,
  pp. 7870--7886, Dec. 2013.

\bibitem{ZhangIT2012}
X.~{Zhang} and P.~H. {Siegel}, ``Adaptive cut generation algorithm for improved
  linear programming decoding of binary linear codes,'' \emph{IEEE Trans. Inf.
  Theory}, vol.~58, no.~10, pp. 6581--6594, Oct. 2012.

\bibitem{Wei2018}
{H. {Wei} and A. H. {Banihashemi}}, ``{An iterative check polytope projection
  algorithm for {ADMM}-based {LP} decoding of {LDPC} codes},'' \emph{{IEEE
  Commun. Lett.}}, vol.~{22}, no.~{1}, pp. {29--32}, {Jan.} {2018}.

\bibitem{7456284}
X.~{Liu} and S.~C. {Draper}, ``The {ADMM} penalized decoder for {LDPC} codes,''
  \emph{IEEE Trans. Inf. Theory}, vol.~62, no.~6, pp. 2966--2984, Jun. 2016.

\bibitem{7740920}
B.~{Wang}, J.~{Mu}, X.~{Jiao}, and Z.~{Wang}, ``Improved penalty functions of
  {ADMM} penalized decoder for {LDPC} codes,'' \emph{IEEE Commun. Lett.},
  vol.~21, no.~2, pp. 234--237, Feb. 2017.

\bibitem{JiaoCL2015}
X.~Jiao, H.~Wei, J.~Mu, and C.~Chen, ``Improved {ADMM} penalized decoder for
  irregular low-density parity-check codes,'' \emph{IEEE Commun. Lett.},
  vol.~19, no.~6, pp. 913--916, Jun. 2015.

\bibitem{8715473}
Y.~{Wei}, M.-M. {Zhao}, M.-J. {Zhao}, M.~{Lei}, and Q.~{Yu}, ``An {AMP}-based
  network with deep residual learning for mm{W}ave beamspace channel
  estimation,'' \emph{IEEE Wireless Commun. Lett.}, vol.~8, no.~4, pp.
  1289--1292, Aug. 2019.

\bibitem{LcgNet2019}
Y.~Wei, M.-M. Zhao, M.-J. Zhao, and M.~Lei, ``Learned conjugate gradient
  descent network for massive {MIMO} detection,'' \emph{arXiv: 1906.03814},
  2019.

\bibitem{7926071}
T.~{Gruber}, S.~{Cammerer}, J.~{Hoydis}, and S.~ten. {Brink}, ``On deep
  learning-based channel decoding,'' in \emph{51st CISS}, Mar. 2017, pp. 1--6.

\bibitem{8242643}
E.~{Nachmani}, E.~{Marciano}, L.~{Lugosch}, W.~J. {Gross}, D.~{Burshtein}, and
  Y.~{Be'ery}, ``Deep learning methods for improved decoding of linear codes,''
  \emph{IEEE J. Sel. Topics Signal Process.}, vol.~12, no.~1, pp. 119--131,
  Feb. 2018.

\bibitem{SparseNN}
W.~Xu, X.~You, C.~Zhang, and Y.~Be'ery, ``Polar decoding on sparse graphs with
  deep learning,'' in \emph{ACSSC}, Oct. 2018, pp. 599--603.

\bibitem{DeepUnfolding1}
J.~R. Hershey, J.~L. Roux, and F.~Weninger, ``Deep unfolding: {M}odel-based
  inspiration of novel deep architectures,'' \emph{arXiv:1409.2574}, Nov. 2014.

\bibitem{DeepUnfolding2}
A.~Balatsoukas-Stimming and C.~Studer, ``Deep unfolding for communications: {A}
  survey and some new directions,'' \emph{arXiv:1906.05774}, Jun. 2019.

\bibitem{Borgerding2016Onsager}
M.~{Borgerding}, P.~{Schniter}, and S.~{Rangan}, ``{AMP-Inspired} deep networks
  for sparse linear inverse problems,'' \emph{IEEE Trans. Signal Process.},
  vol.~65, no.~16, pp. 4293--4308, Aug. 2017.

\bibitem{Yang2016Deep}
Y.~Yang, J.~Sun, H.~Li, and Z.~Xu, ``Deep {ADMM-Net} for compressive sensing
  {MRI},'' in \emph{NIPS}, 2016, pp. 10--18.

\bibitem{ADAM}
D.~P. Kingma and J.~Ba, ``Adam: {A} method for stochastic optimization,'' in
  \emph{ICLR}, 2015.

\bibitem{channelcodes}
M.~Helmling, S.~Scholl, F.~Gensheimer, T.~Dietz, K.~Kraft, S.~Ruzika, and
  N.~Wehn, ``{D}atabase of channel codes and {ML} simulation results,''
  \url{www.uni-kl.de/channel-codes}, 2017.

\bibitem{crvali1995}
R.~Kohavi, ``A study of cross-validation and bootstrap for accuracy estimation
  and model selection,'' in \emph{IJCAI}, 1995, pp. 1137--1143.

\end{thebibliography}
\bibliographystyle{IEEETran}

%








\end{document}